\documentclass[a4paper,fleqn,usenatbib]{mnras}

\usepackage{mathptmx}
\usepackage{txfonts}

\usepackage[T1]{fontenc}
\usepackage{ae,aecompl}

\usepackage{graphicx}	
\usepackage{url}        
\usepackage{multirow}   
\usepackage{amssymb}	
\usepackage{subfig}     
\usepackage{textcomp}   
\usepackage{caption}    
\usepackage{hyperref}   

\newcommand{\Hersc}{{\it Herschel}}
\newcommand{\hevics}{HeViCS}

\newcommand{\IRAS}{{\it IRAS\/}}
\newcommand{\ISO}{{\it ISO\/}}
\newcommand{\Spitzer}{{\it Spitzer\/}}

\newcommand\dg{$^\circ$}

\title[Far-reaching Dust Distribution in Galaxy Disks]{Far-reaching Dust Distribution in Galaxy Disks}

\author[M. W. L. Smith et al.]{Matthew W. L. Smith,$^{1}$\thanks{Contact e-mail: \href{mailto:Matthew.Smith@astro.cf.ac.uk}{Matthew.Smith@astro.cf.ac.uk}}
Stephen A. Eales$^{1}$, Ilse De Looze$^{2,3}$, Maarten Baes$^{2}$,\newauthor
George J. Bendo$^{4}$, Simone Bianchi$^{5}$, M\'{e}d\'{e}ric Boquien$^{3,6}$, Alessandro Boselli$^{7}$,\newauthor
Veronique Buat$^{7}$, Laure Ciesla$^{8,9}$, Marcel Clemens$^{10}$, David L. Clements$^{11}$,\newauthor
Asantha R. Cooray$^{12}$, Luca Cortese$^{13}$, Jonathan I. Davies$^{1}$, Jacopo Fritz$^{2,14}$,\newauthor
Haley L. Gomez$^{1}$, Thomas M. Hughes$^{2,15}$, Oskar \L. Karczewski$^{16}$, Nanyao Lu$^{17}$,\newauthor
Seb  J. Oliver$^{16}$, Aur\'{e}lie Remy-Ruyer$^{18}$, Luigi Spinoglio$^{19}$, and Sebastien Viaene$^{2}$.
\\
$^{1}$School of Physics \& Astronomy, Cardiff University, The Parade, Cardiff, CF24 3AA, UK.\\
$^{2}$Sterrenkundig Observatorium, Universiteit Gent, Krijgslaan 281 S9, B-9000 Gent, Belgium.\\
$^{3}$Institute of Astronomy, University of Cambridge, Madingley Road, Cambridge, CB3 0HA, UK.\\
$^{4}$Jodrell Bank Centre for Astrophysics, Alan Turing Building, School of Physics and Astronomy,\\
\ The University of Manchester, Oxford Road, Manchester, M13 9PL, UK.\\
$^{5}$INAF-Osservatorio Astrofisico di Arcetri, Largo Enrico Fermi 5, 50125 Firenze, Italy.\\
$^{6}$Unidad de Astronom\'{i}a, Fac. de Ciencias B\'{a}sicas, Universidad de Antofagasta, Avda. U. de Antofagasta 02800, Antofagasta, Chile.\\
$^{7}$Laboratoire d'Astrophysique de Marseille-LAM, Universit\'{e} d'Aix-Marseille \& CNRS, UMR7326, 38 Rue F. Joliot-Curie, F-13388,\\ 
\ Marseille Cedex 13, France.\\
$^{8}$University of Crete, Department of Physics, Heraklion 71003, Greece.\\
$^{9}$Institute for Astronomy, Astrophysics, Space Applications and Remote Sensing, National Observatory of Athens, GR-15236 Penteli, Greece.\\
$^{10}$INAF-Osservatorio Astronomico di Padova, Vicolo dell'Osservatorio 5, 35122 Padova, Italy.\\
$^{11}$Astrophysics Group, Imperial College, Blackett Laboratory, Prince Consort Road, London SW7 2AZ, UK.\\
$^{12}$Center for Cosmology and the Department of Physics \& Astronomy, University of California, Irvine, CA 92697, USA.\\
$^{13}$International Centre for Radio Astronomy Research, University of Western Australia, 35 Stirling Highway, Crawley, WA 6009, Australia\\
$^{14}$Centro de Radioastronom\'\i a y Astrof\'\i sica, UNAM, Campus Morelia, A.P. 3-72, C.P. 58089, Mexico.\\
$^{15}$Instituto de F\'isica y Astronom\'{i}a, Universidad de Valpara\'{i}so, Avda. Gran Breta\~{n}a 1111, Valpara\'{i}so, Chile.\\
$^{16}$Astronomy Centre, Department of Physics and Astronomy, University of Sussex, Brighton, BN1 9QH.\\
$^{17}$Infrared Processing and Analysis Center, California Institute of Technology, MS 100-22, Pasadena, CA 91125, USA.\\
$^{18}$CEA, Laboratoire AIM, Irfu/SAp, Orme des Merisiers, F-91191, Gif-sur-Yvette, France.\\
$^{19}$Istituto di Fisica dello Spazio Interplanetario, INAF, Via del Fosso del Cavaliere 100, I-00133 Roma, Italy.}

\date{Accepted 2016 July 4. Received 2016 July 4; in original form 2015 December 6}

\pubyear{2016}

\begin{document}
\label{firstpage}
\pagerange{\pageref{firstpage}--\pageref{lastpage}}
\maketitle

\begin{abstract}
  In most studies of dust in galaxies, dust is only detected from its emission to approximately the optical radius of the galaxy. 
  By combining the signal of 110 spiral galaxies observed as part of the Herschel Reference Survey, we are able to improve our
  sensitivity by an order-of-magnitude over that for a single object. Here we report the direct detection of dust from
  its emission that extends out to at least twice the optical radius. We find that the distribution of dust is consistent with an
  exponential at all radii with a gradient of $\sim -1.7$\,dex\,$R_{25}^{-1}$. Our dust temperature declines linearly from $\sim$25\,K 
  in the centre to 15\,K at $R_{25}$ from where it remains constant out to $\sim$2.0\,$R_{25}$. 
  The surface-density of dust declines with radius at a similar rate to the surface-density of stars but more slowly than the surface-density
  of the star-formation rate. Studies based on dust extinction and reddening of high-redshift 
  quasars have concluded that there are substantial amounts of dust in intergalactic space. By combining our
  results with the number counts and angular correlation function from the SDSS, we show that with 
  Milky Way type dust we can explain the reddening of the quasars
  by the dust within galactic disks alone. Given the uncertainties in the properties of any intergalactic dust,
  we cannot rule out its existence, 
  but our results show that statistical investigations of the dust in galactic halos that use the reddening
  of high-redshift objects must take account of the dust in galactic disks.
\end{abstract}




\section{Introduction}

Recent studies of the extinction of quasars \citep{Menard2010} and the reddening of galaxies \citep{Peek2015}
imply that approximately 50\% of the interstellar dust in the universe lies outside galactic disks. However,
we still know remarkably little about how far out the dust in galactic disks extends.

Spiral galaxies have huge disks of atomic hydrogen that extend much further
than the standard optical radius \citep[][$R_{25}$, the radius where the 
optical brightness corresponds to a \textit{B}-band brightness of 25\,mag\,arcsec$^{-2}$]{Bigiel2012}, but the atomic hydrogen
is unprocessed material that may have fallen on the galaxies from outside. 
Studies into the extent of molecular gas using the carbon monoxide (CO) line, even  when combining the results
from a very large number of galaxies, have only managed to trace the molecular gas out to 1.1\,$R_{25}$ \citep{Schruba2011}.

There have been many attempts to detect dust around nearby galaxies by using both its absorption and reflection properties in the optical, 
and its emission properties in the far-infrared. Observations of edge-on galaxies have been used to look for dust above the
plane of the galaxy. \citet{Hodges-Kluck2014} found that UV emission extends between 5--20\,kpc from the mid-plane of the galaxies in their sample, 
and was consistent with the emission being light from the disk scattered by dust above the disk. They found that the halo dust has an exponential 
form with scale heights $\sim 2.5-5$\,kpc, much larger than typical scale heights of stellar thick disks \citep[0.3--1.5\,kpc][]{Hodges-Kluck2014}. 
\citet{Bocchio2015} combined \Hersc\ and \Spitzer\ data for the edge-on galaxy NGC\,891, finding extended emission above the disk
with a scale height of 1.44\,kpc.

The extent of dust disks has been investigated by looking for the effects of dust extinction of background sources
(e.g., quasars, background galaxies) through the outskirts of a nearby galaxy. \citet{Holwerda2009} have shown in a pair of occulting
galaxies that dust extinction can be reliably detected out to 1.5\,$R_{25}$. 
Attempts to detect dust in the outer disk from its emission were undertaken with the early infrared space telescopes, \IRAS\ and \ISO. However, they
were limited by sensitivity, wavelength coverage and resolution. \citet{Nelson1998} obtained a 2$\sigma$ detection 
at 1.5\,$R_{25}$ using \IRAS\ by a stacking analysis, and \citet{Alton1998} 
used \ISO\ observations of a sample of eight nearby galaxies to show
that the dust has a larger exponential scale-length than the optical emission.
Recently \Spitzer\ and \Hersc\ data have been used to create radial profiles of the dust for the nearby galaxies in the SINGS/KINGFISH sample \citep{Munoz-Mateos2009b, Hunt2015}. 
The KINGFISH radial profiles presented in \citet{Hunt2015} typically extend to between 1.0--1.5\,$R_{25}$ at 250\micron, across a variety of morphological types.

In this paper, we aim to go beyond the typical detection radius of dust for an individual galaxy, by combining the signal from a large sample of galaxies. Since the dust in the 
outskirts of galaxies is likely to be very cool ($\leq 15$\,K) with low surface brightness, 
the \textit{Herschel Space Observatory} \citep{Pilbratt2010} is the 
only telescope with enough sensitivity to have the potential to directly detect this emission. We have selected galaxies from the Herschel Reference Survey (HRS) \citep{Boselli2010}, the largest targeted survey of nearby
galaxies with \Hersc. These galaxies all have either spiral or irregular morphology and lie face-on, giving us the best possible
physical resolution. In Section~\ref{sec:data} we present the data used for this analysis. Section~\ref{sec:Method} describes our method for creating the averaged radial profiles of the dust
and other galaxy components. Section~\ref{sec:results} presents our results, and the conclusions are presented
in Section~\ref{sec:conc}.

\section{The Sample}
\label{sec:data}

The Herschel Reference Survey (HRS) is a volume-limited sample of 322 galaxies with distances between 15 and 25\,Mpc \citep{Boselli2010}. To ensure
a reasonable detection rate with \Hersc, the HRS also has a 
2MASS K-band magnitude limit of $K_{Stot} \leq 12$ for late-type galaxies.
Observations with the SPIRE instrument \citep{Griffin2010} at 250, 350 and 500\micron\ and with the PACS instrument \citep{Poglitsch2010} at
100 and 160\micron\ exist for all the galaxies.
Eighty three objects fall within the Virgo cluster and as such have been observed by the 
Herschel Virgo Cluster Survey (\hevics) \citep{Davies2010, Davies2012}. For all late-type galaxies the HRS observations were designed to cover an
area of at least 1.5 times the optical diameter with a minimum size of 4\arcmin $\times$ 4\arcmin. 
In practice, the SPIRE images generally cover a much larger area around each galaxy, allowing
us to trace the dust to larger radii. Each observation of a late-type galaxy has an instrumental 1$\sigma$ sensitivity of 5.64, 5.65,
and 6.60\,mJy/beam (0.57, 0.32, and 0.18\,MJy/sr) for the 250, 350 and 500\micron\ bands, respectively. The instrumental noise is comparable 
to the confusion noise of 5.8, 6.3, and 6.8\,mJy/beam \citep{Nguyen2010} for the 250, 350 and 500\micron\ bands, respectively. 
Full details about the survey can be found 
in \citet{Boselli2010}.

\section{Methodology}
\label{sec:Method}

\subsection{Radial Profiles and Averaging}
\label{sec:process}

To obtain the highest sensitivity to extended dust we average the emission from a large sample of HRS galaxies to create an average
surface-brightness profile. We included galaxies in the sample if they met the
following criteria:
\begin{enumerate}
  \item The galaxy morphology must be classified as a spiral or irregular. The classifications are the same as used in other HRS studies \citep{Cortese2012}.
	    Ellipticals and S0s were not included as they are known to contain much less dust and have
		much smaller dust disks than spirals and irregulars \citep{Smith2012a}.
  \item The galaxy cannot have an inclination greater than 60\dg. The inclinations were calculated from the major and minor diameters
	    measured by the HRS team \citep{Cortese2012b} using images from the Sloan Digital Sky Survey.
  \item The galaxy must have been detected by \Hersc\ with emission that is clearly extended \citep{Ciesla2012}.
  \item There must not be another extended \Hersc\ source that could affect the measured radial profile 
        (i.e., the sources must be separated by a distance of at least 6.0\,$R_{25}$).
  \item The \Hersc\ image must not have significant contamination from diffuse emission from Galactic interstellar dust.
  \item The image of each galaxy must extend to a distance of at least 2.5\,R$_{25}$ from the centre of the galaxy to make it possible to remove
        any residual background emission that is not from the galaxy itself.
\end{enumerate}
Once all the criteria had been applied, there were 117 galaxies, providing an order of 
magnitude increase in sensitivity over what is possible for a single object.

We obtained a surface-brightness profile for each galaxy by calculating the distance to each
pixel from the centre of the galaxy and then expressing this distance as a fraction of 
the optical radius ($R_{25}$). We calculated this distance using the coordinates of the pixel, the optical radius, 
the inclination (estimated from the eccentricity), the distance to the galaxy \citep{Cortese2012} and the optical centre. 
A line of fixed radius would appear as an ellipse on a SPIRE map.
We extended the profile out until a radial annulus reaches the edge of the image.
Before calculating the average profile for the whole sample, we removed a constant
offset from each galaxy's profile using the mean value of 
the surface brightness between 2.2 $\leq R/R_{25} <$ 2.7 (a couple of profiles do not extend as far as 2.7\,$R_{25}$, but contain enough
pixels for an accurate background subtraction). This choice of background subtraction means by construction we are only
sensitive to dust within 2.2\,$R_{25}$.
We then placed all the pixels from all images into radial bins and calculated the 
mean value in each bin and its error ($\sigma / \sqrt{N}$, where N is the number of data points and $\sigma$ is the standard deviation).
To ensure we account for all uncertainties (e.g., background variations), we also ran a simulation by 
carrying out an identical analysis using the same number of positions and the same radial bins as in our 
real analysis at random positions
on the \hevics\ map (but avoiding Virgo cluster objects). We then performed the same stacking analysis, 
and re-ran the process a 1000 times. By measuring the standard deviations
in each bin, we get a robust estimate for the minimum error in each radial bin. We then use
this uncertainty if it is larger than the error above.

\begin{figure*}
  \centering
  \includegraphics[trim=15mm 16mm 6mm 5mm,clip=True, width=0.89\textwidth]{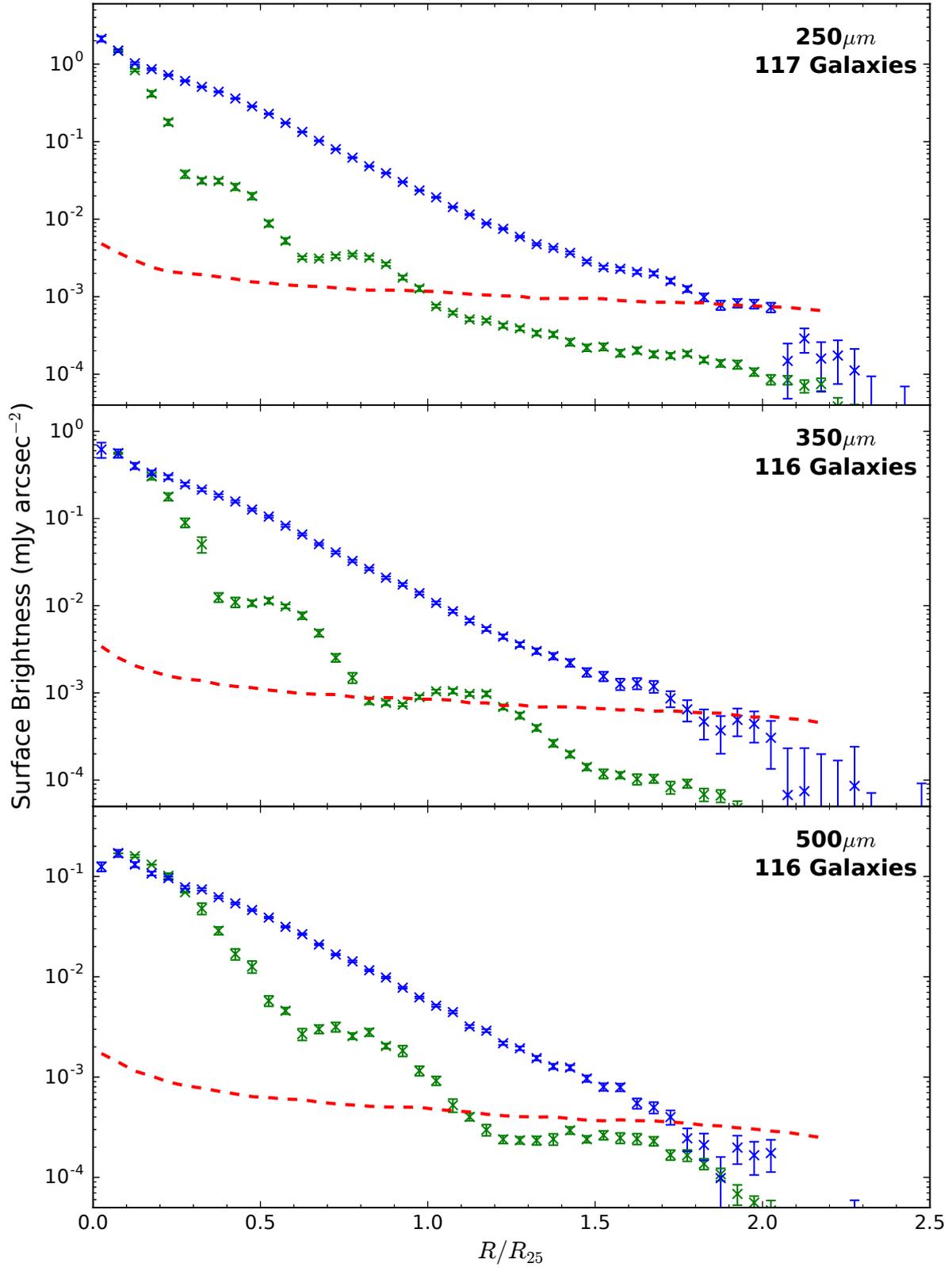}
  \caption{The average emission from dust plotted against the radial distance  
           for the late-type galaxies in the Herschel Reference Survey. Before averaging, the radial distance has
           been normalised for each galaxy by the optical radius ($R_{25}$) of the galaxy.  		   
  		   The blue data points show the average radial profile
           for each wavelength. The wavelength and number of galaxies that has been used in each plot are shown in the top-right
           corner. The green data points represent the profile of a point source processed using the same method and scaled by
           the median value of $R_{25}$ for the sample. The dashed red line shows the 3$\sigma$ sensitivity limit found from simulations using
           the background regions of the Herschel Virgo Cluster Survey map.}
  \label{fig:aveProfile}
\end{figure*}

The results are shown in Figure~\ref{fig:aveProfile}. The red-dashed lines in Figure~\ref{fig:aveProfile} show the 3$\sigma$ detection level found from the background simulations described in the previous paragraph. 
The median optical radius for the sample is 1.25\arcmin\ ($R_{25}$ ranges from 0.33 to 4.83\arcmin) and the median distance is
17.0\,Mpc. Therefore, the median value of $R_{25}$ corresponds to a physical radius of 6.2\,kpc.

\begin{figure*}
  \centering
  \includegraphics[trim=15mm 16mm 6mm 5mm,clip=True, width=0.89\textwidth]{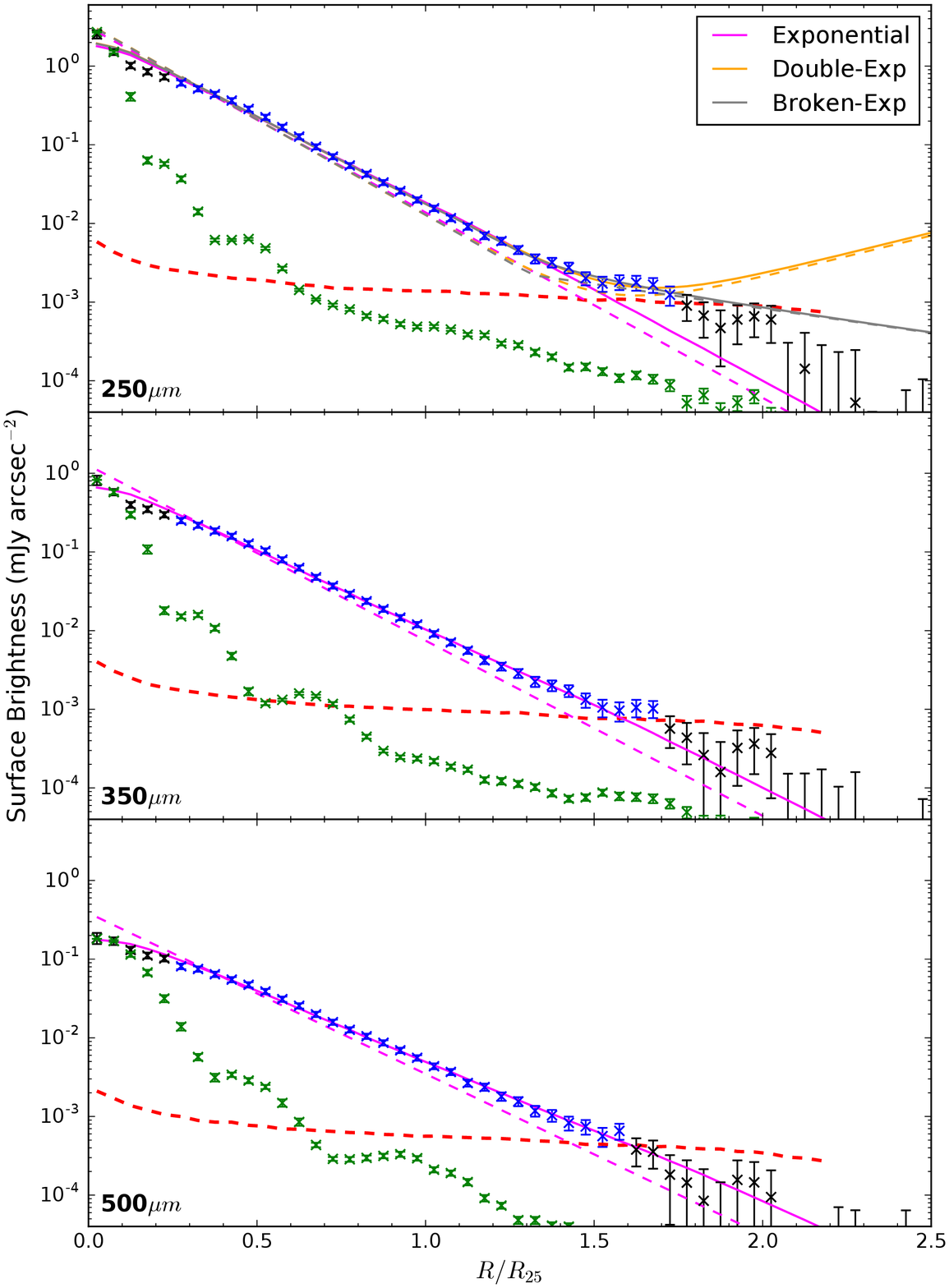}
  \caption{The best-fit models to the measured average surface-brightness profiles of the sample of 
           large galaxies (45 objects). The curved red dashed line shows the 3$\sigma$ limit (Section \ref{sec:process})
           with the blue points showing the measurements that are above the limit. As in Figure \ref{fig:aveProfile}, the green data points
           represent the profile of a point source, processed using the same method and scaled by the median value of $R_{25}$ for the 
           sample of large galaxies,. The three models are
           described in Section \ref{sec:modFit} and are only fit to the blue data points. The colour key for
           the three models is shown in the legend. Only one model is shown at 350 and 500\micron\ because the other more complex 
           models did not provide a significantly better fit to the data (see description in Section\,\ref{sec:modFit}).
           The dashed and the solid lines show the models before and
           after they have been convolved with the PSF.}
  \label{fig:deconModels}
\end{figure*}

\subsection{Reliability}
\label{sec:reliability}

To check that our average profile is representative of the sample of galaxies and not just dominated by a small subset of objects, we performed
a monte-carlo test. We randomly split the sample into 4 sub-sets and measured a profile  for each subset as described above.
For each profile we measured the radius where the surface-brightness falls below 0.002\,mJy\,arcsec$^{-2}$ at 250\micron. We repeated this
process 10000 times and found no sub-set where the profile is significantly steeper than the one we show in Figure~\ref{fig:aveProfile}, 
showing our result is statistically robust. 

An important question is whether at large radii, where the surface-brightness is very 
low, we are effectively measuring the point spread function (PSF) of the observations. To test this, we used
SPIRE calibration observations of Neptune. We found the average surface-brightness profile of the point spread function
using the same procedure as for the galaxies, scaling it to the same physical units as the galaxies
using the median $R_{25}$ of the sample.
The Neptune profile has been normalised to the second data point of the 
averaged galaxy profile and is shown in green in Figure \ref{fig:aveProfile}.
In the figure we see that in all bands the average galaxy profile is significantly higher at all radii than
the expected flux from a point source. The shallow slope of the galaxy profile compared to the PSF precludes the possibility
that dust detected at large $R$ is from the Airy rings of the brighter inner parts of the galaxy.
Using the median $R_{25}$ to scale the
Neptune profile probably over-estimates the effect of the PSF as the galaxies 
with larger angular sizes contribute more pixels to the average profile.
 
\begin{figure*}
  \centering
  \includegraphics[trim=0mm 0mm 0mm 0mm,clip=True, width=0.95\textwidth]{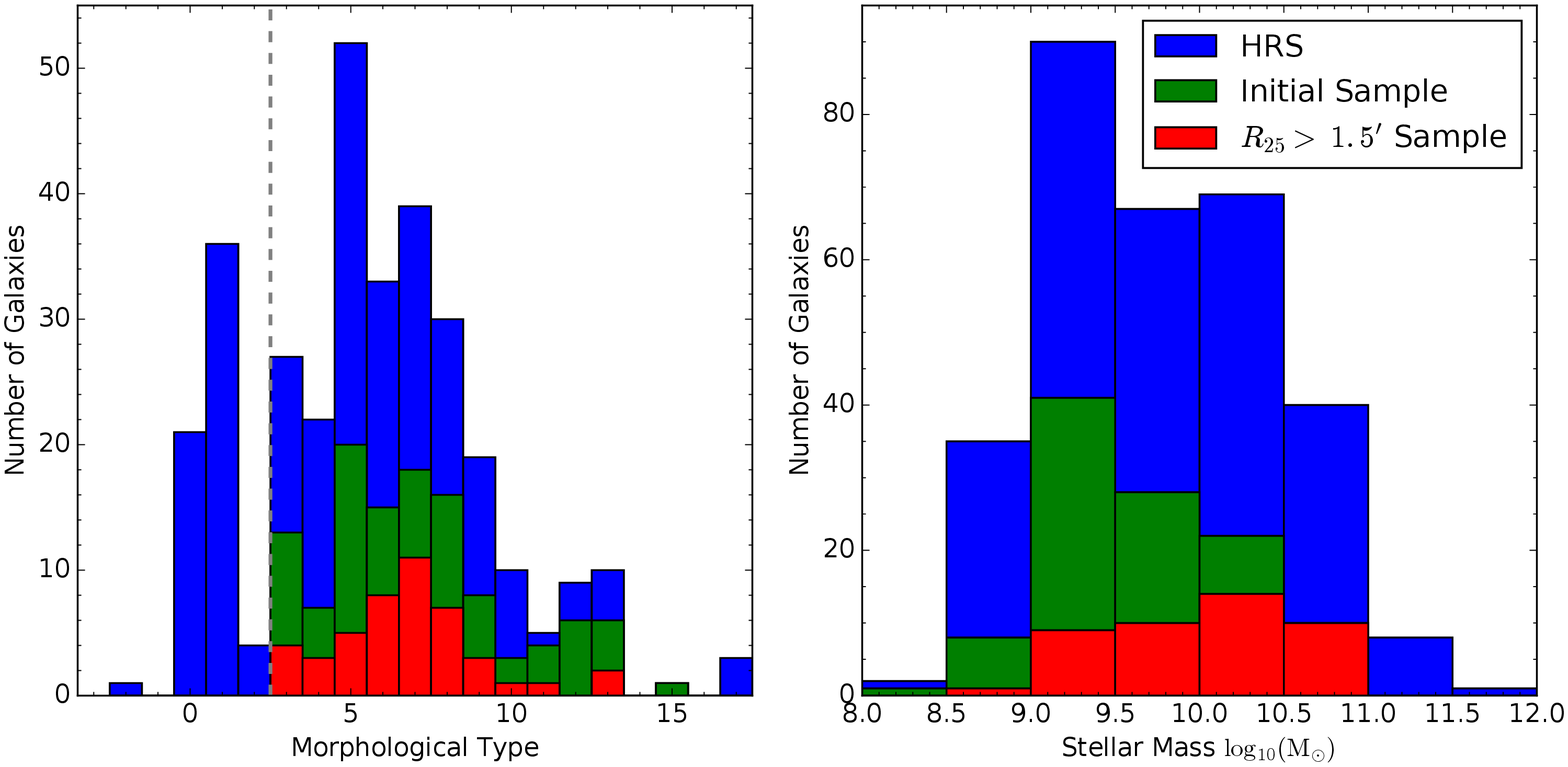}
  \caption{The distributions of morphology and stellar mass for the entire HRS (blue), our initial sample of 117 late-type galaxies (green) and the
           sample of 45 galaxies with $R_{25} > 1.5^{\prime}$ (red). The morphological classification is the Goldmine \citep{Gavazzi2003} classification system,
           where types 3--10 represent Sa to Sdm galaxies, 11--14 are irregular/peculiar galaxies and 14--17 are BCD's. Morphological types -3--2 represent early-type
           galaxies which are not used in this study; the grey-dashed line shows the boundary between early-type and late-type galaxies. The stellar masses
           are taken from \citep{Cortese2012b}. Note that 10 galaxies of the HRS do not have stellar mass determinations, with 7 and 1 of these in the initial and
           large sample, respectively, and so are not included in the plot.}
  \label{fig:sampleProp}
\end{figure*}

As a final check that we are not simply detecting at large radii the wings of the point spread function,
we repeated the analysis for the 45 galaxies with $R_{25} > 1.5$\arcmin (the largest is 4.83\arcmin). We find a much
larger difference between the average surface-brightness profile and the wings of the point spread function 
(Fig.~\ref{fig:deconModels}). Figure~\ref{fig:sampleProp} shows the morphological-type and stellar mass distributions
for the entire HRS, our initial sample of 117 galaxies, and our sample of galaxies with $R_{25} > 1.5$\arcmin. 
Our initial sample of 117 galaxies has a similar distribution to the entire HRS for both morphological type (excluding early-types) and stellar mass, but our sample with $R_{25} > 1.5$\arcmin\ is biased towards higher stellar masses. 
As mentioned in Section \ref{sec:data} 83 of our galaxies fall within the Virgo cluster, 
and so the galaxies may be deficient in dust because they are deficient in
H{\sc i} \citep{Cortese2010}. However, we saw little difference in the results when we split the
sample into classes based on measurements of the H{\sc i} deficiency.
To minimise the effect of the PSF, we used this sample for the rest of the analysis described in this paper.
The median optical radius for this sample is 2.1\arcmin\ and the median distance is
17.0\,Mpc. Therefore, the median value of $R_{25}$ corresponds to a physical radius of 10.3\,kpc.

\subsection{Model Fitting Procedure}
\label{sec:modFit}

To fit analytic models to the measured average surface-brightness profiles in Figure~\ref{fig:deconModels}, 
we need to take into account the point spread function, which is large at submillimetre 
wavelengths.
We fitted three different models to the 250--500\micron\ emission:
\begin{itemize}
  \item A single exponential. This has two free parameters: a scale length and the normalization of the function.
  \item Double exponential (sum of two exponentials). This has four free parameters: the two normalisations and scale
		length of each function.
  \item Broken exponential (an exponential where at a given radius the scale-length changes). This has four free parameters:
the overall normalisation, the scale length of each exponential, and the radius dividing the regions fitted by the
two functions.
\end{itemize}
The fitting procedure was similar to a standard profile fitting procedure with a chi-squared minimisation except that 
we performed a full 2D-convolution by projecting our model onto an
image, convolving the model image with the PSF, creating a model radial profile and then comparing it to the observed radial profile. To decide
what is the best model, we compared the reduced chi-squared value of the three models. 
The fits were performed over the region in which there is at least a 3$\sigma$ detection at 250\micron, except we did not fit the model
to the central two data points (equivalent to central 30\arcsec) to ensure there is no effect from the nuclei. 
We found that the broken-exponential model provided a significantly better fit at 250\micron\ but at 350 and 500\micron\ there
was no statistical reason to prefer the more complex models over the single exponential (Fig.~\ref{fig:deconModels}).
Virtually all iterations of the monte-carlo test described in Section \ref{sec:reliability} show evidence for the shallower
slope at large radii suggesting we are not being influenced by a small subset of galaxies.

Before deriving dust temperatures, we first convolved all images to the same resolution as the SPIRE 500\micron\ images
which have the lowest angular resolution of all our images \citep[using the method described in][]{Aniano2011}.
We then calculated empirical average radial profiles in the SPIRE bands for the 45 HRS galaxies
with $R_{25} > 1.5$\arcmin, using the same method as above. Dust temperatures tend to increase towards
the centres of galaxies \citep{Smith2010,Smith2012b,Foyle2012,Mentuch2012}, which means that data at shorter
wavelengths than the SPIRE wavelengths is necessary for accurate temperature measurements. We determined 
average surface-brightness profiles at 100 and 160\micron\ from the images of the HRS galaxies made with
the PACS camera \citep{Poglitsch2010, Cortese2014} on \Hersc.
However, when estimating temperatures we only used
PACS points at $R<0.5R_{25}$, since background variations in the PACS images
appear to dominate at larger radii. While only using three data points to find a temperature is not ideal, 
at $R>0.5$\,$R_{25}$ the dust is colder and so it becomes easier to estimate a dust temperature
from the three SPIRE measurements alone than at smaller radii.

\begin{figure*}
  \centering
	  \includegraphics[trim=0mm 0mm 0mm 0mm,clip=True, width=0.95\textwidth]{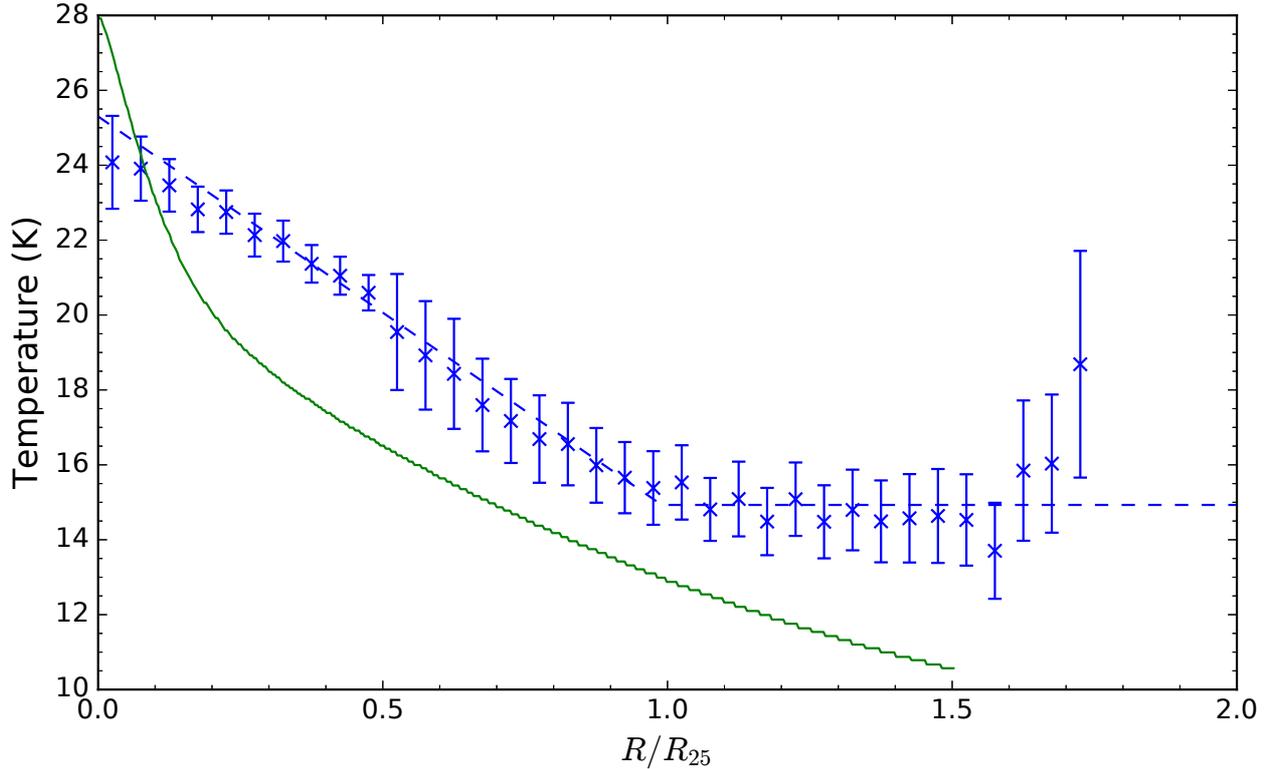}
  	  \caption{The relationship between dust temperature and galactocentric radius 
  	           from fitting a modified blackbody to the spectral energy distribution of the dust emission (blue points). 
			   The dashed line shows a model in which the dust temperature declines linearly with radius 
			   until it reaches a minimum.	           
  	           The green line shows the prediction of a model of how the dust in a disk is heated by starlight 
  	           (Sect. \ref{sec:dustRadPro}).}
 	 \label{fig:tempModels}
\end{figure*}

We estimated dust temperatures by fitting a modified blackbody to the five or three flux measurements for each radial bin,
taking full account of the correlated flux errors between the different \Hersc\ bands \citep{Smith2012b}. 
As we were limited to only three data points for $R>0.5R_{25}$ we had to assume a constant dust-emissivity index ($\beta$). We chose
to use $\beta = 2$ which is the typical value found in resolved studies of nearby galaxies \citep{Smith2012b,Eales2012,Kirkpatrick2014,Hughes2014}. 
The uncertainties on the dust temperature were calculated by a Monte-Carlo approach; in each radial bin every data point
was re-sampled using its value and uncertainty (taken correct account of the part of the calibration uncertainty that
is correlated between bands) and the modified blackbody
fit performed again. From the distribution of temperature
values obtained from the 1000 runs of the Monte-Carlo simulation, we estimated the uncertainty in the dust temperature (typically $\sim$2\,K).
Our temperature profile is shown in Figure \ref{fig:tempModels}.
We fitted a simple model to the data points, where the dust temperature decreases linearly with radius until it reaches a minimum.
We then derived the dust surface density by combining our best-fit model for the 
dust emission and our model of the dust temperature. We assume a value of
0.192\,$\rm m^2$\,$\rm kg^{-1}$ at 350\micron\ \citep{Draine2003} for the dust absorption coefficient ($\kappa_{\nu}$).

This value is chosen to match the values used in other HRS and \hevics\ papers, and is based on Milky Way type dust.
For this work we have assumed that the value of $\kappa$ is constant out to a radius of 2.0\,$R_{25}$,
as metallicity gradients in galaxies tend to be flat \citep{Moustakas2010, Sanchez2014} and that we are measuring dust in the disks of galaxies. However, as very little is known about dust in galaxy outskirts it is possible that the value of $\kappa$ 
could have a significant systematic change with radius.
The absolute value of $\kappa_{\nu}$ is also an important consideration when we compare our results 
with optical reddening studies and so we discuss the assumptions and uncertainties of $\kappa{_\nu}$ 
in Section~\ref{sec:halos}.

\begin{figure*}
  \centering
  	\includegraphics[trim=0mm 0mm 0mm 0mm,clip=True, width=0.95\textwidth]{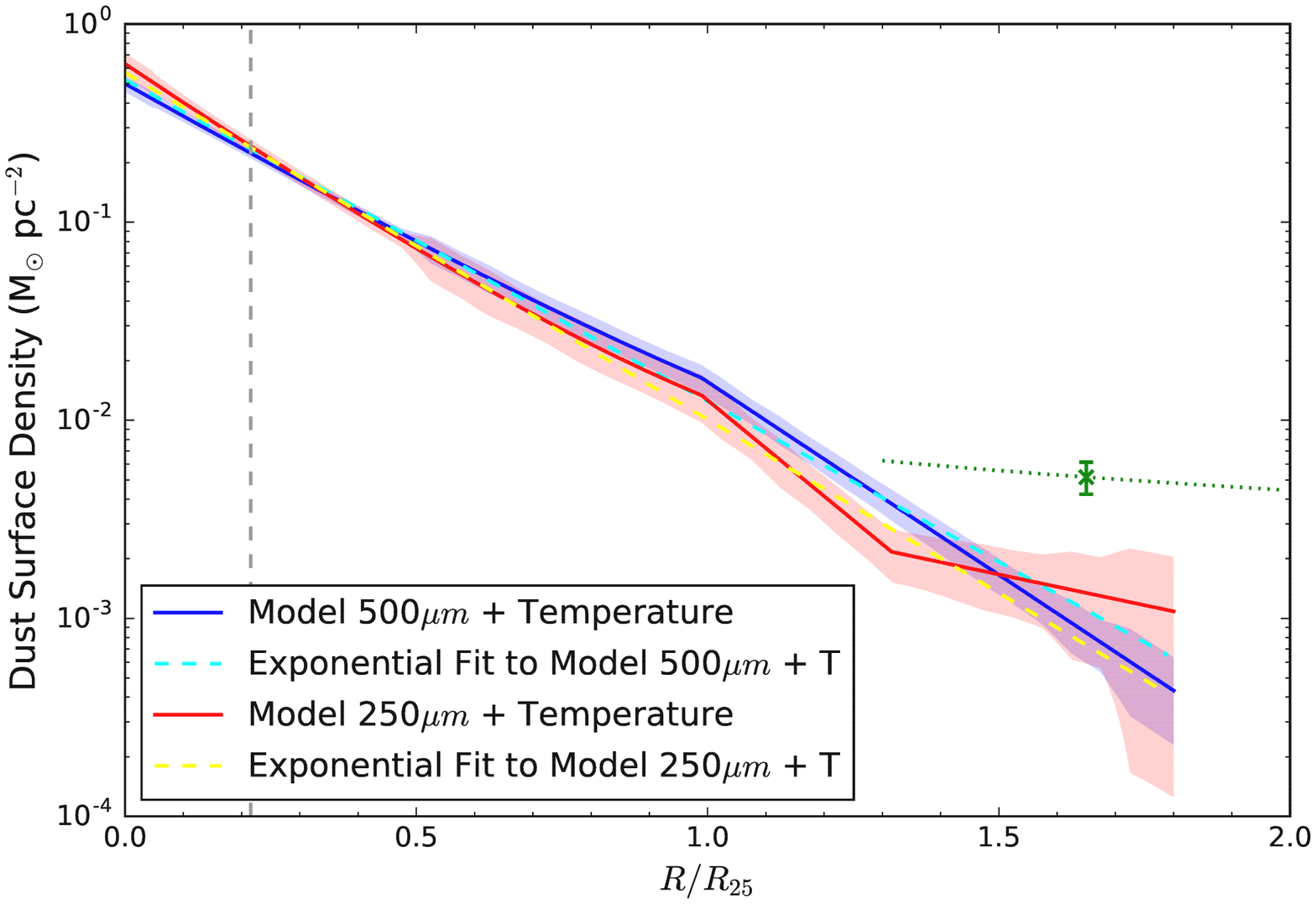}
  	\caption{The surface density of dust plotted against radius, obtained by combining our fitted dust emission models 
  	         with the temperature model shown in Figure~\ref{fig:tempModels}. The solid lines show our best estimates 
  	         of the dust surface-density for our best-fitting dust emission models at 250 and 500\micron, 
             with the shaded regions showing the 1$\sigma$ uncertainties. The dashed cyan and yellow lines are the best-fit 
             exponential models to the empirical dust surface-density distributions. The grey vertical dashed line shows 
             the radius below which the data points are not fit to ensure there is no effect from galactic nuclei. 
             The green data point is the one overlapping data point from the quasar reddening study of \citet{Menard2010}, 
             and the green dotted line is their estimate of the relationship between dust surface density and radius, 
             which varies with radius as $r^{-0.8}$ (green line). If \citet{Menard2010} does represent the true distribution
             of dust our disagreement in surface densities may be the result of the range of radii that contribute to the  
             \citet{Menard2010} data point or our background subtraction.}
  	\label{fig:dustMassProfile}
\end{figure*}

Figure~\ref{fig:dustMassProfile} shows the results. The two solid lines show the surface-density of dust we derive when starting from
the best-fit models to the 250\micron\ and 500\micron\ surface-brightness profiles. The dashed lines show fits of
single-exponential models to these empirical dust column-density distributions.

\subsection{Radial Variation of Other Galaxy Phases}
\label{sec:other}

This work requires comparing our dust profile to the other constituents which make up galaxies. 
For the molecular gas and total gas we used the results available in literature \citep{Schruba2011,Bigiel2012}, 
derived from the HERACLES \citep{Leroy2009} and THINGS \citep{Walter2008} samples. 
This could introduce a potential bias due to the different selection criteria used for HERACLES and THINGS 
relative to the HRS, but equivalent data is not available for the HRS sample. 

To derive the radial distributions of stellar surface-density and star-formation rate (SFR) surface-density,
we used two alternative datasets. \Spitzer\ and GALEX data exists for many of the galaxies (see Table~\ref{tab:scaleFactors}), and we used this data
to determine empirical average radial distributions of stellar surface-density and SFR surface-density to compare
with the radial distributions of the other galaxy components. However, since 24\micron\ \Spitzer\ data does
not exist for the whole sample and a few of the 3.6\micron\ images are smaller than ideal, we also estimated profiles using WISE data, which does exist for the entire sample.
In the case of the WISE data, we first convolved all the images to the same resolution as the 
250\micron\ image to make the comparison with the dust surface-density as fair as possible.
We did not do this when using the \Spitzer\ data as the higher-resolution profile provides an independent check of 
the model fitting applied to the \Hersc\ data.
Table~\ref{tab:scaleFactors} summarises the datasets.
We obtained very similar results whichever dataset we started with.

To find the radial variation of the surface-density of the stars, we assumed the 3.6\micron\ emission is 
proportional to the column-density of the stars. We used \Spitzer\ 3.6\micron\ observations which are available for
44 of the 45 in our sample of large galaxies and were obtained as part of the S$^4$G survey \citep{Sheth2010}. 
Our second dataset was the WISE 3.4\micron\ data which exists for all 45 galaxies in the sample. 
In the case of this dataset, we first convolved the images to the same resolution as the 250\micron\ images before estimating
the average radial distribution of stellar surface density.

To find the radial variation of the star-formation rate surface-density, we first created a SFR map for each galaxy by combining 
an obscured star-formation tracer (either 22 or 24\micron\ emission) and an un-obscured star formation tracer, 
the ultraviolet emission, using
published recipes \citep{Leroy2008}. Ideally we would have used the FUV data from GALEX, but many of the observations are too 
shallow for this analysis. Instead, we used GALEX NUV images collected for the HRS \citep{Cortese2012b} that exist for all
but one of our sample, and assumed a conversion factor of 0.657
to convert NUV fluxes to FUV fluxes (the conversion factor was calculated from the global 
UV fluxes for the sample of large galaxies). To estimate the obscured component, we used the \Spitzer\ 24\micron\ data \citep{Bendo2012}, 
which exists for about half the sample, and WISE 22\micron\ data, which exists for the whole sample. 
In the case of the WISE data, we first convolved the WISE and NUV images to the same resolution as the 250\micron\ images
before producing the SFR maps for each galaxy.
We also tested the effect of applying a small correction, using the 3.6\micron\ emission, to 
account for the stellar contribution to both the infrared and ultraviolet data \citep{Leroy2008}. This correction was found to have a negligible effect on our 
conclusions. To summarise, we were able to create high-resolution SFR maps (resolution of $\sim$5\arcsec) for 28 of our 45 galaxies and maps at similar 
resolution to the SPIRE 250\micron\ images for 44 out of 45 galaxies. 

Before producing the average distributions of either stellar surface-density or SFR surface-density, 
we masked any bright objects on the images, which is especially important due to
contamination from stars in these wavebands, and also removed a linear background from the images.
In creating the distributions we used very similar methods to the one used to create the radial distributions
of submillimetre emission. The smaller sizes of the \Spitzer\ 3.6\micron\ and 24\micron\ images meant that
our default background annulus was reduced to 2.0--2.5\,$R_{25}$.
However, for about half the galaxies the 3.6\micron\ images were still too small for the background to be determined from this annulus.  
For these galaxies, we instead estimated the background on each image from an annulus with 1.0\,$< R/R_{25} <$\,1.5 and then added a 
corrective offset to the profile calculated in the same region from the average profile for the other galaxies.

\begin{figure*}
  \centering
  \includegraphics[trim=0mm 0mm 0mm 0mm,clip=True, width=0.95\textwidth]{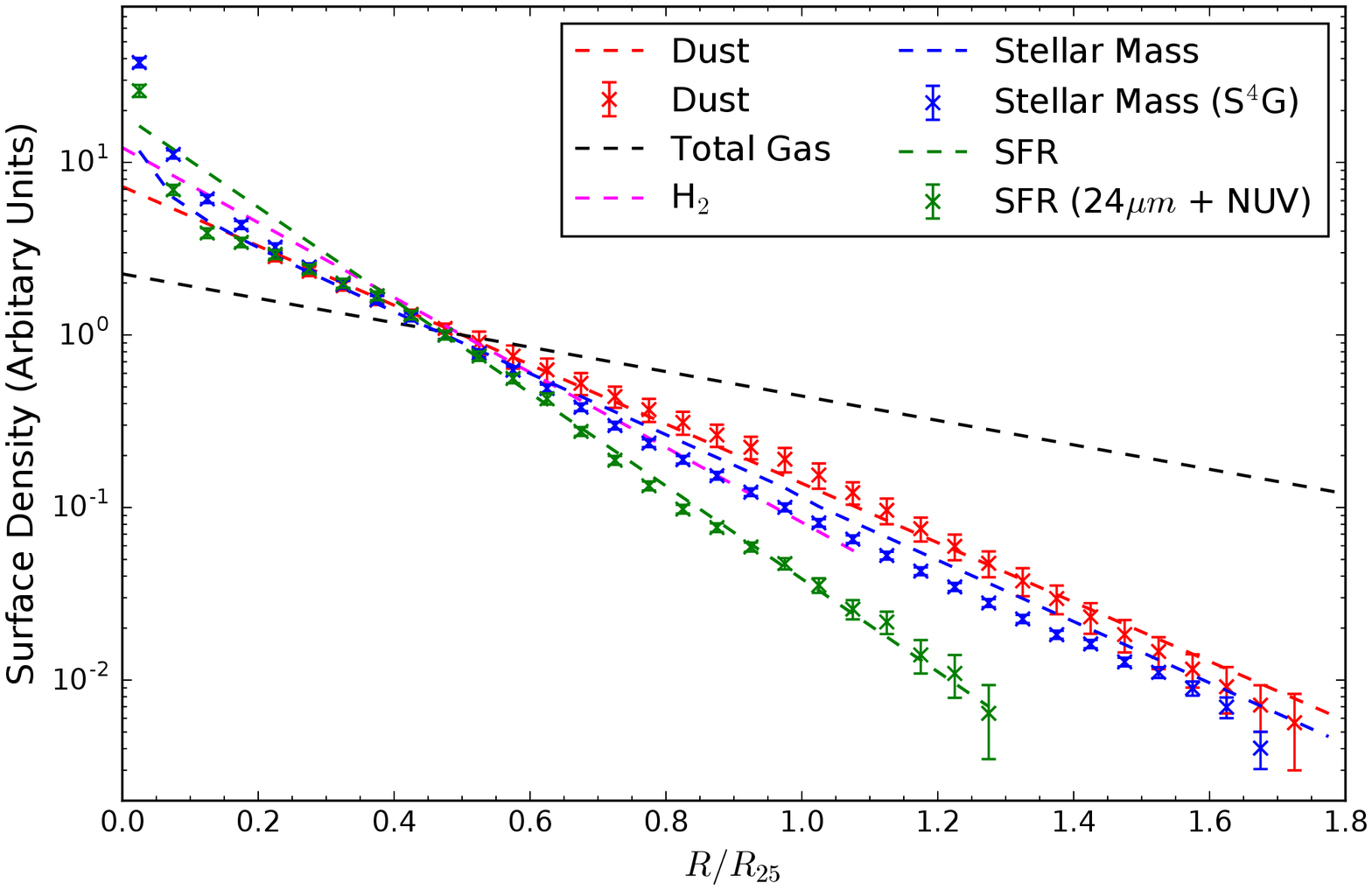}
  \caption{Average surface-density verses radius for all components of a galaxy. The data points show 
  		   the estimated average surface-density for the HRS galaxies in each radial bin for dust (red)
  		   star-formation rate (green) and stellar mass (blue). The stellar mass and SFR data points
  		   are radial profiles measured from high resolution datasets ($\leq$ 6.0\arcsec FWHM) from \Spitzer\
  		   and GALEX. The dashed lines of the same colour show the best-fit exponential models for 
  		   images which have been convolved to the same resolution as our 250\micron\ images  
  		   and have had the same procedure applied to produce the average profiles. In this case
  		   we have used WISE data instead of \Spitzer\ data as observations exist for the entire sample and cover a larger area.
  		   In the case of the total gas (black) and molecular gas (magenta) we have shown the exponential distributions 
  		   derived from samples of local galaxies in the literature (Sect. \ref{sec:other}).}
  \label{fig:multiBand}
\end{figure*}

\begin{table*}
\caption{Exponential Scale Factors}
\begin{tabular}{lcccc}
\hline \hline
Component & Data   & Numbers   & Matched     &  Gradient\\
          & Source & of Objects & Processing  & (dex $R_{25}^{-1}$)\\
\hline
Dust         & (250\micron\ \& T)     & 45 & Y & -1.72\,$\pm$\,0.01\\
Dust         & (500\micron\ \& T)     & 45 & Y & -1.75\,$\pm$\,0.02\\
Stellar Mass & WISE, 3.4\micron        & 45 & Y & -1.78\,$\pm$\,0.07\\
Stellar Mass & S$^4$G, 3.6\micron      & 44 & N & -1.90\,$\pm$\,0.03\\
SFR          & WISE 22\micron\ \& NUV & 44 & Y & -2.69\,$\pm$\,0.05\\
SFR          & MIPS 24\micron\ \& NUV & 28 & N & -2.58\,$\pm$\,0.05\\
Total Gas\,$^{\rm a}$    & H{\sc i} \& $\rm H_2$  & 33  & - & -0.71\\
$\rm H_2$\,$^{\rm a}$    & -                      & 33  & - & -2.17\\
Metallicity\,$^{\rm a}$  & -                      & 306  & - & -0.16 (with $\sigma = 0.12$)\\
\hline
\end{tabular}\\
\begin{flushleft}
\textbf{Notes.} The gradients from exponential fits to data of different components of a galaxy. The `Matched Processing' column lists whether the data have been
convolved to the 250\micron\ resolution (or already have similar resolution) and have had the same procedure applied as for the dust profiles, or if the model 
has been fit to high-resolution data. Components marked $^{\rm a}$ are gradients taken from the literature \citep{Bigiel2012,Schruba2011,Sanchez2014} for
different samples of galaxies.
\end{flushleft}
\label{tab:scaleFactors}
\end{table*}

\section{Results \& Discussion}
\label{sec:results}

\subsection{Dust Radial Profile}
\label{sec:dustRadPro}

The surface-brightness profile produced from combining the radial profiles from the 117 images (see Sect.~\ref{sec:process}) obtained with the \Hersc\ SPIRE camera)
is shown in Figure \ref{fig:aveProfile}. Emission from the dust is detected at $>3\sigma$ to $R=1.975$\,$R_{25}$ at 250\micron\ and 
at $>3\sigma$ to $R=1.75$\,$R_{25}$ at 350 and 500\micron\ (the $3\sigma$ sensitivity limit is shown by the red-dashed line in the figure).
While an increase to $R$\,$\sim$2.0\,$R_{25}$ may not sound a significant improvement over studies that traced dust to $R$\,$\sim$1.2\,$R_{25}$, this corresponds to over a factor of ten
increase in sensitivity.

The average surface-brightness profile
for the sample of large galaxies and the best fit models to the profile are shown in Figure~\ref{fig:deconModels}. 
We find that at 250\micron\ the average surface-brightness profile is best fit with a broken
exponential, but that at 350 and 500\micron\ a single exponential is sufficient (for details of the models see Sect.~\ref{sec:modFit}).

The continuum emission from dust depends both on the column-density and temperature 
of the dust, and so to determine the column density, we determined the average line-of-sight temperature of the dust using the method described in \ref{sec:modFit}.
On the assumption that the dust is heated by starlight, both the light from the old stellar population and the hot young stars in star-formation regions,
we have used our stellar surface-density, SFR surface-density 
and dust-mass profiles to predict the dust temperature. We used the Stellar 
Kinematics Including Radiative Transfer (SKIRT) code \citep{Camps2015}, which 
models the radiative transfer of starlight through dust using a monte-carlo based approach. 
The results of the temperature prediction are shown in Figure~\ref{fig:tempModels}.

The best-fit temperature (Fig.~\ref{fig:tempModels}) declines linearly 
from $\simeq$24\,K in the centre to approximately 15\,K at  a radius of $\approx R_{25}$ where the temperature
appears to reach a minimum. Our radiative transfer model of the dust in a galaxy predicts a similar 
decline in the dust temperature with radius but does not predict this plateau in the 
dust temperature (Fig.~\ref{fig:tempModels}). We note that fairly high dust temperatures
well outside disks have been seen before, with \citet{Bocchio2015} finding that the dust around the
edge-on galaxy NGC\,891 still has a temperature of $\sim$19\,K 4\,kpc above the plane.

\citet{Kashiwagi2015} estimated that the cosmic UV background \citep{Xu2005}
would produce a minimum dust temperature of $\sim$10\,K.
However, given dust radiates with a $T^6$ dependence,
the UV background would have to be higher by $\sim 1.5^6$ to explain our minimum temperature of 15\,K.
Another possible heating source is from collisional heating in a hot X-ray gas.
\citet{Bocchio2014} found dust that had been pulled out of NGC\,4338 was heated to 20--30\,K in the hot X-ray gas in the Virgo Cluster, and \citet{Yamada2005} 
show that dust could reach $\sim$30\,K in the centres of clusters. However, although spiral galaxies are known to have halos of hot gas
\citep[e.g., ][]{Anderson2015} it is unclear whether this gas could heat the dust to $\sim$15\,K. 

We used our temperature estimates and the models that fit the surface-brightness
profiles at the SPIRE wavelengths to determine the radial variation in the column-density
of the dust (Fig.~\ref{fig:dustMassProfile}). We find that the dust column-density has an exponential decline with radius 
(gradient $\sim$-1.7\,dex\,$R_{25}^{-1}$), no matter which model of the dust emission we start with. Throughout this paper 
the gradients ($\alpha$) are given in units of dex\,$R^{-1}_{25}$ 
(i.e., a value of -1 corresponds to a decrease by a factor of 10 over a radial distance of $R_{25}$,
and can be described by:
\begin{equation}
\Sigma = \Sigma_{0} 10^{\alpha r}
\label{equ:dustDist}
\end{equation}
where $\Sigma$ is the surface density, $\Sigma_{0}$ is the surface density when $r=0$, and $r$ is the radius in angular units.
If we extrapolate our exponential surface-density model to an infinite radius, we estimate that the 
dust between $1.0$\,$R_{25} < R \leq \infty$ constitutes only 9.3\% of all the total dust in the disk

\subsection{Comparison With Other Galaxy Components}

Figure~\ref{fig:multiBand} shows all galaxy components plotted on one diagram and Table~\ref{tab:scaleFactors} 
gives the gradients of the exponential models that best fit the radial distributions of stellar surface-density,
SFR surface-density, dust surface density, molecular gas and total gas.
In the case of stellar surface-density and 
SFR surface-density, we obtained very similar values for the gradient whether we started with the high-resolution 
images or the images convolved to the same resolution as the 250\micron\ images.
The blue and green points in Figure~\ref{fig:multiBand} show the empirical average radial distributions of stellar surface-density and
SFR surface-density estimated from the high-resolution datasets. The blue and green dashed lines show the fits of 
single exponential models to the empirical radial distributions of stellar surface-density and SFR surface density 
derived from the low-resolution datasets. In this case, the empirical distributions are not shown in Figure~\ref{fig:multiBand}.

As expected, the average surface-density of all galaxy components falls with distance from the 
centre of the galaxy. As expected, given the extended distribution
of atomic gas around galaxies \citep{Leroy2008}, the total column-density of gas declines much more slowly than the other 
components. The gradients of the distribution of stellar mass and of dust are in good agreement, but the star-formation
rate falls more rapidly with radius than either of the other two.

\citet{Hunt2015} also find similar distributions of the dust and stars, finding the scalelengths of KINGFISH spirals to have a value of 
$R_{250}/R_{25} \approx 0.37$ at 250\micron\ versus $R_{3.6}/R_{25} \approx 0.35$ at 3.6\micron\ (1.17\,dex\,$R_{25}^{-1}$ and 1.24\,dex $R_{25}^{-1}$, 
respectively, in our gradient units). 
Our results show that the similarity between the distributions of dust and stellar mass extends out to $R$\,$\sim$\,1.8\,$R_{25}$.
Our dust gradient of -1.75\,dex\,$R_{25}^{-1}$ is steeper than the value of 1.17\,dex\,$R_{25}^{-1}$ for the KINGFISH spiral galaxies \citep{Hunt2015},
although the KINGFISH dust distributions are distributions of dust 250\micron\ emission rather than dust surface-density.
\citet{Munoz-Mateos2009b} find the median 
gradient of the radial distribution of the surface density of dust for the entire SINGS survey of 1.49\,dex\,$R_{25}^{-1}$, significantly closer to our value. 

It is still uncertain whether dust is mostly formed in evolved stars \citep{Matsuura2009,Boyer2012}, supernovae \citep{Gomez2012,Indebetouw2014} or 
formed by grain growth in the ISM \citep{Ossenkopf1993,Hirashita2011}. 
We note that if dust grains are formed in supernovae and then fairly quickly destroyed, we would expect 
the surface-densities of the dust and the star-formation rate to have the same dependence on radius, which is not found. 

\subsection{Dust, Gas and Metallicity}

Metallicity gradients in galaxies tend to be small. The 
average metallicity gradient for a sample of 306 galaxies measured with CALIFA is -0.16\,dex\,$R_{25}^{-1}$ \citep[scatter of $\sim$0.12\,dex\,$R_{25}^{-1}$][]{Sanchez2014}, from
observations of H{\sc ii} regions out to 1.3\,$R_{25}$ (there is possible evidence of a flattening at larger radii). 
For the galaxies in the SINGS survey the measured radial metallicity gradients are steeper,
$\sim$-0.42\,--\,-0.33\,dex\,$R_{25}^{-1}$ \citep[depending on the calibration used][]{Moustakas2010}. On the assumption that
these metallicity measurements apply to all phases of the interstellar medium, we combine the gradient in metallicity from the CALIFA measurement and the gradient in
the column-density of all gas (Table~\ref{tab:scaleFactors}) and find that the gradient in the column-density of metals is 
$\sim$-0.9\,dex\,$R_{25}^{-1}$, again significantly flatter than the dust gradient of -1.7\,dex\,$R_{25}^{-1}$.
Similar radial declines in the dust-to-metal ratio have been seen in M31 \citep{Mattsson2014} and in SINGS galaxies \citep{Mattsson2012b}.
In closed-box chemical evolution models this radial decline can be explained by dust grain growth in the ISM \citep{Mattsson2012a}.

\subsection{Are Galaxies Surrounded by Extended Halos of Dust?}
\label{sec:halos}

Based on the angular correlation between the reddening of $\sim$85,000 high-redshift
quasars and the positions of $\sim$2.4\,$\times$\,$10^7$ galaxies detected in the Sloan
Digital Sky Survey \citep[SDSS ][]{York2000}, \citet[][henceforth M10]{Menard2010} concluded that galaxies
are surrounded by very extended halos of dust. They found that the mean dust surface-density varies
with angular distance, $\theta$, from the position of an SDSS galaxy as
\begin{equation}
\Sigma_d(\theta) = 2.5 \times 10^{-3} \left({\theta \over 0.1\ {\rm arcmin} }\right)^{-0.8}\ h\ {(M_{\odot}\ pc^{-2})}
\label{equ:Menard}
\end{equation}
(taken from Figure 8 of M10), with this distribution extending to a physical distance of several Mpc ($h$ is $H_{0}/100$). \citet{Kashiwagi2015}
have suggested that the reddening measurements are actually due, not to extended halos of dust around individual galaxies, but
to the dust within galactic disks and the fact that the positions of galaxies are correlated. Their evidence for this claim are the
results from stacking measurements around the positions of the same SDSS galaxies, using the far-infrared image of the sky made with \IRAS.
They, too, detect extended dust around the SDSS galaxies, but this time in emission, and by comparing the reddening and emission measurements,
they estimate the temperature of the dust is $\simeq$20\,K, which they argue is what one would expect for dust in the galactic disks rather than dust well
outside galactic disks (however, see Sect. \ref{sec:dustRadPro}).

The strong practical motive for trying to distinguish between these two possibilities is the need to understand the effect of 
dust on experiments to measure cosmological parameters, such as experiments that use the magnitudes of Type Ia supernovae 
to measure $\Omega_{\Lambda}$ and the equation-of-state of dark energy \citep{Menard2010b}. If all galaxies are surrounded by such
extended halos of dust, all measurements of the magnitudes of high-redshift objects need to be corrected for the effects of dust extinction and reddening. But, if the the conclusion of \citet{Kashiwagi2015} is correct, and almost all dust is confined to galaxy disks, only a small
percentage of lines-of-sight actually pass through galactic disks (we quantify this below). We now discuss what our results can contribute to
resolving the controversy. We borrow the nomenclature of \citet{Kashiwagi2015} in which the first model is referred to as the 
\textit{circumgalactic dust model} (CGD model) and the latter model as the \textit{interstellar dust model} (ISD model).

Our result that the dust column-density of galaxies declines more rapidly with radius out to $R$\,$\simeq$\,2\,$R_{25}$
than given by equation~\ref{equ:Menard} (Figure~\ref{fig:dustMassProfile}) does not allow us to discriminate between the ISD and the CGD models. This is because equation~\ref{equ:Menard} is only valid beyond the minimum radius for which M10 have a reddening measurement which 
is only slightly less than 2\,$R_{25}$ (Figure~\ref{fig:dustMassProfile}), where we are not sensitive. 
Therefore it is possible that the surface-density of dust declines rapidly with radius out to 
$R$\,$\simeq$\,2\,$R_{25}$ but then follows the more gentle relationship given by equation~\ref{equ:Menard}.

We now adopt the alternative approach of combining the statistical properties of the SDSS galaxies---their number counts and 
clustering properties---with the average dust surface-density radial profile for the HRS galaxies (Figure~\ref{fig:dustMassProfile}) to
test whether we can produce a relationship similar to equation~\ref{equ:Menard} from the dust within galaxy disks.

M10 used SDSS galaxies with $i < 21$, whereas our study is ultimately based on the photometric $B$-band, since $R_{25}$ is the 
distance from the centre of a galaxy out to an isophote with a surface brightness in the $B$-band of 25\,mag\,arcsec$^{-2}$. To use
a sample of galaxies as similar as possible to that used by M10 and to connect the sample to the standard definition of optical radius
we used the galaxy number counts derived from the SDSS $r$-band images and an average $B-r$ colour of 1.65 \citep{Yasuda2001}.
Since the mean $r-i$ colour measured by \citet{Yasuda2001} is $\simeq$0.4, we used the number counts with $r < 21.5$.

Of course, not all galaxies detected in the SDSS contain much dust, and in particular early-type galaxies
contain small amounts of dust \citep{Smith2012a}. \citet{Eales2015} have recently measured the percentage of the stellar mass
in the nearby Universe that is in late-type galaxies is $\sim$49\%, using a definition of `late-type' that means a late-type galaxy is almost
certainly dominated by a disk. Informed by this result, we multiply the SDSS number counts by 0.5 to produce an estimate 
of the number counts of SDSS galaxies which contain a disk, $N_{disk}$. We then make the assumption that each of these
galaxies has the dust surface-density radial profile shown in Figure~\ref{fig:dustMassProfile}. Although these are large assumptions,
they are not as important as they appear because of the larger uncertainty in the amount of evolution of dust in galaxies, which we discuss below.

On the assumption that the the surface brightness of a galaxy depends on radius with the form shown in equation~\ref{equ:dustDist}, the total flux
density out to the standard optical radius is given by
\begin{equation}
I_{tot} = {2 \pi \, I_{25} \, \over {\left( \alpha\,{\rm log_e(10)} \right) ^2}}  \left( \left(\alpha\,R_{25}\, {\rm log_e(10)} - 1 \right) + 10^{-\alpha\,R_{25}} \right)
\label{equ:totFlux}
\end{equation}
in which $I_{25}$ is the surface brightness corresponding to 25 mag\,arcsec$^{-2}$. Because $\alpha$
is given in units of $R_{25}$ (Table \ref{tab:scaleFactors}), this equation makes it possible to calculate $\theta_{25}$,
and thus the area subtended on the sky, for a galaxy of any apparent magnitude.
\citet{Pohlen2006} have fit exponential
models to the SDSS images of 90 face-on late-type galaxies. The mean value of $\alpha$ derived from the values given by them for the
inner scale length in the $g$-band for each galaxy (their Table~3) is $-1.64 \pm 0.08$\,dex\,$R_{25}^{-1}$. Given the average value of $\alpha$ we can use
equation~\ref{equ:totFlux} to calculate the optical radius of a galaxy of any apparent magnitude.

A very simple thing we can then do is calculate the fraction of the sky covered by the disks. As we can only detect dust out to a radius
of 2\,$\times$\,$R_{25}$, we assume that a disk only extends out to this radius. Using this assumption, the number counts for late-type
galaxies, $N_{disk}$, and the average value of $\alpha$ from \citet{Pohlen2006}, we estimate that the percentage of the sky covered by 
disks is $0.44 \pm 0.05$\%, with the errors obtained using the error for the mean value of $\alpha$ given above. This low percentage
confirms the importance of distinguishing between the CGD and ISD models for making the correct corrections for dust \textit{extinction} and 
reddening when carrying out cosmological experiments, because in the ISD model a very small fraction of the sky is covered by dust.

For a galaxy in which the dust surface-density varies with radius with the relationship given in equation~\ref{equ:dustDist} the
mean column density out to twice the optical radius (2\,$\times$\,$\theta_{25}$) is given by:
\begin{equation}
<\Sigma_{d,galaxy}> = {\Sigma_0 \left( 10^{2\alpha\,R_{25}} \left( 2 \alpha\,R_{25}\, {\rm log_e(10)} - 1 \right) + 1\right)\over {2 \left( \alpha\,R_{25}\,{\rm log_e(10)} \right) ^2}} 
\end{equation}
We use the values for $\Sigma_0$ and $\alpha$ for the mean dust surface-density profile shown in Figure~\ref{fig:dustMassProfile}:
$\Sigma_0 = \rm 0.6$\,$M_{\odot}$\,$pc^{-2}$ and $\alpha=-1.75$\,dex\,$R_{25}^{-1}$.
The expected average dust column-density profile around
SDSS galaxies is then given by:
\begin{equation}
\Sigma_d(\theta) = \int_0^{21} <\Sigma_{d,galaxy}> 4 \pi\ (\theta_{25})^2\ w(\theta,m)\ N_{disk}(m)\ dm
\end{equation}
In this equation, $w(\theta,m)$ is the angular correlation function for SDSS galaxies. \citet{Connolly2002} have measured the angular correlation
function for SDSS galaxies as a function of $r$ magnitude. They found that at all magnitudes the angular correlation function is well represented 
by $w(\theta) = A \theta^{-\gamma}$, with $\gamma \simeq 0.7$ and $\rm log_{10}A$ a linear function of $r$-band magnitude. In calculating 
$w(\theta,m)$, we assume the same value of $\gamma$ and estimate $A$ at each magnitude from a linear fit to the data shown in Figure~3 of \citet{Connolly2002}.

\begin{figure*}
  \centering
  \includegraphics[trim=0mm 0mm 0mm 0mm,clip=True, width=0.95\textwidth]{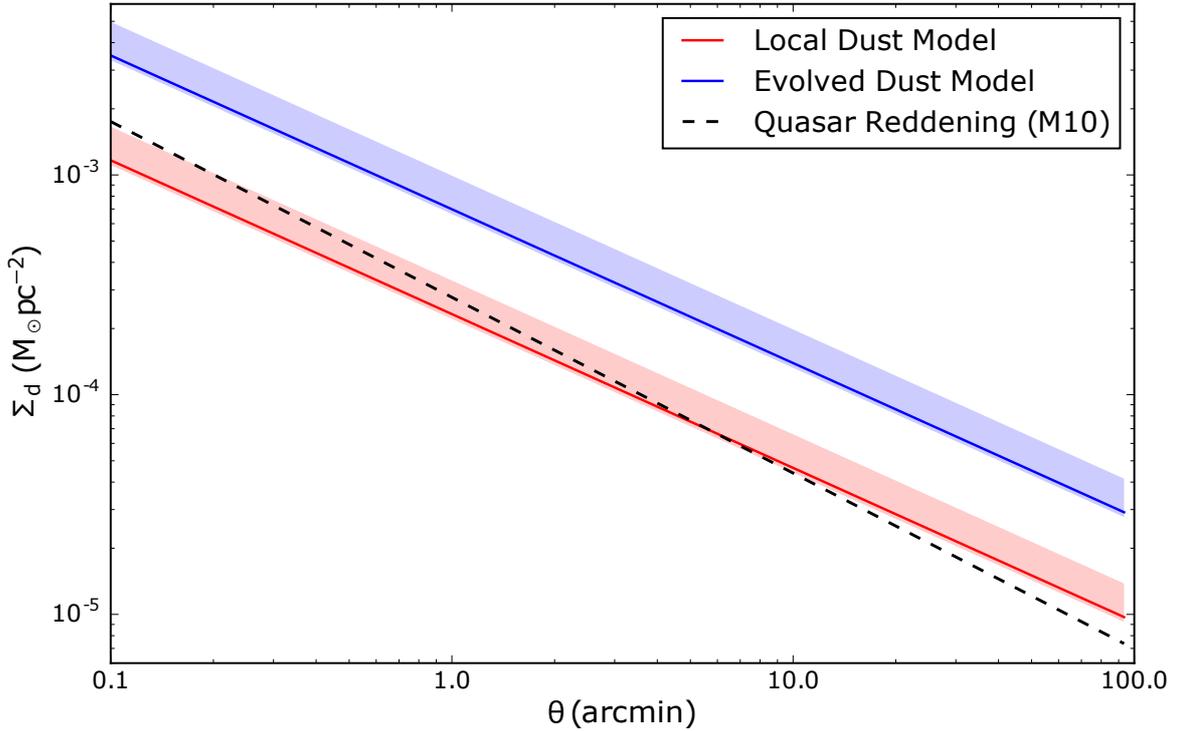}
  \caption{The relationship between mean dust surface-density and angular distance from the centre of an SDSS galaxy. The dashed black line 
  shows the relationship estimated by M10 from looking at the correlation between the reddening of high-redshift quasars and the positions of SDSS galaxies.
  The solid red line shows our prediction for this relationship considering only the dust in galactic disks, with the shaded region showing the predictions
  if we use values for $\alpha$ in our model between the upper and lower 1$\sigma$ errors in the mean value of $\alpha$ (see text for details). The blue line show our
  prediction if we incorporate the evolution in the dust masses of galaxies measured by \citet{Bourne2012}.}
  \label{fig:halos}
\end{figure*}

Figure~\ref{fig:halos} shows our estimate of $\Sigma_d(\theta)$ from this procedure. The red solid line in the figure shows our prediction with the mean value 
of $\alpha$ we calculated from the results in \citet{Pohlen2006} and the shaded region show the predictions if we use values for $\alpha$ between  
the $\pm$1\,$\sigma$ errors on the mean. The black dashed line shows the relationship measured by M10 (equation~\ref{equ:Menard}). This relationship
is very similar to the result from the reddening study (M10).

This is without considering the effect of galactic evolution, as the mean redshift of the SDSS galaxies is $\simeq$0.3. 
Two results, both from the \Hersc-ATLAS survey \citep{Eales2010}, have shown that there
is substantial evolution in the amount of dust in galaxies by this redshift. \citet{Dunne2011} have shown that the dust masses of galaxies
have increased by a factor of $\sim$4.6 by this redshift. This result is based on the small fraction of galaxies with very high dust masses, but
\citet{Bourne2012} have carried out a stacking analysis on a large sample of SDSS galaxies which is sensitive to galaxies with much lower dust masses,
finding that the average dust mass of optically-selected blue galaxies  of the same stellar mass 
has increased by a factor of $\simeq$3 by a redshift of 0.3. In Figure~\ref{fig:halos}
we have shown our estimate of $\Sigma_d(\theta)$ scaled by a factor of 3.0 to allow for the effect of galaxy evolution.
This now exceeds the relationship found in the reddening study.

An important consideration is the absolute and relative uncertainties in the values of $\kappa_\nu$ used for both our analysis of 
dust emission and the reddening study of M10. Our value of $\kappa_{\nu} = 0.192$\,$\rm m^2$\,$\rm kg^{-1}$ at 350\micron\ \citep{Draine2003} 
assumes Milky-Way like carbonaceous and silicate dust, and is in the typical range of values
used of $\kappa_{\nu}$ in local galaxies studies \citep[see][and references there in]{Clark2016}. Our $\kappa_\nu$ value
is higher by a factor of $\sim$two than the latest empirical value derived in this study \citep{Clark2016}; using this value
would roughly double the amount of dust we measure in the disks of galaxies. As we are only
measuring the dust emission in the disks of galaxies these values should be reasonable values to assume out to 2.0\,$R_{25}$,
particularly as the metallicity gradients of galaxies tends to be small. A greater unknown is what the extinction coefficient would
be for dust in the IGM; M10 assume an SMC-type dust due to the lack of a 0.2\micron\ bump in low-ionisation
absorbers in the halo's of $\sim L^\star$ galaxies \citep[e.g.,][]{York2006} and the
fact that extinction curves for high-redshift galaxies also do not show the 0.2\micron\ bump. 
If larger Milky-Way-type dust were assumed, the halo dust surface density
implied by the M10 results would approximately double. Of course, the dust in intergalactic space 
might be very unlike the dust within galactic disks.
Sputtering is likely to be important \citep{Yamada2005} and possibly only large grains would
survive being expelled from galactic disks
\citep{Croft2000}. We emphasise, however, that in our model 90\% of the dust contributing to the average surface
density shown in Figure~\ref{fig:halos} is produced by dust within a radius less than $R_{25}$, where our knowledge of
the properties of dust is much better.

Of course there are also large uncertainties in our model, including the division between early-type and late-types, 
the assumption that all late-type galaxies have the same distribution of dust, and the amount of galactic evolution. 
Because of these uncertainties and because of the huge uncertainty as to whether the dust in intergalactic 
space is like dust in galactic disks, we can not rule out the CGD model. All we can say is that
with some assumptions about the properties of dust (e.g., Milky Way type dust grains), the results of M10 might be explained 
by the dust within individual galaxies rather than very extended halos 
of dust around all galaxies.

Despite using results from most sensitive ever far-infrared space telescope, and then combining the results
in a powerful statistical stacking analysis, we have still only managed to detect the continuum emission from
dust out to a distance of twice $R_{25}$ from the galactic centre. Therefore the only practical way to
investigate the properties of dust at greater distances from centres of galaxies
is still through statistical investigations of the reddening of high-redshift objects.
Over three decades, investigations of the dust associated with damped Ly\,$\alpha$ quasar absorption-line systems
have produced many significant detections of this dust and have even produced
detailed measurements of the properties of this dust
\citep[e.g.,][]{Fall1989, Khare2004, Vladilo2008, Wild2007, York2006}.
There have also been detections through quasar absorption-line of gas and metals in the halos of intervening individual galaxies, 
with projected distances of up to 100\,kpc \citep[e.g.,][]{Steidel2002, Chen2005}, although some galaxies fail to show any
absorption \citep{Bechtold1992}. 
Investigations of the correlation of quasar absorption-line systems
with galaxy redshift surveys imply that the absorption-line gas
is often a long way from the centre of a galaxy 
(\citealp{Churchill2005} and references therein; \citealp{Perez-Rafols2015}),
which suggests that much of the dust detected via reddening of quasars is also outside galactic disks. 
The method used by M10 is an elegant way of both measuring the properties of the
dust and also its distribution around galaxies. We have shown that there could be a significant contribution to the reddening
measured by M10 from dust within galactic disks.

Fortunately, there is actually a simple way
of adapting the M10 technique to allow for this contamination.
Even if the CGD and ISD models predict very similar relationships for how the mean reddening depends on distance from an SDSS
galaxy, in the ISD case this relationship is caused by the small percentage of lines of sight that pass through galactic disks, 
whereas in the CGD case it is caused by all lines of sight. Therefore, 
a simple way to distinguish between the two hypotheses would be to measure the variance of the quasar 
reddening measurements, which should be much higher for the ISD model than the CGD model - a test that 
should be easy to carry out using the measurements described by M10.

\section{Conclusions}
\label{sec:conc}

In this paper we present the results of combining images of a large sample of nearby galaxies made from observations with the 
\textit{Herschel Space Observatory}. By combining the images, we determine how the average surface-density of dust depends on
radius. We find the following results:
\begin{enumerate}

\item We detect the emission of dust at $>$\,3$\sigma$ threshold out to a radius of 2.0\,$R_{25}$ at 250\micron\ and
out to a radius of 1.75\,$R_{25}$ at 350 and 500\micron, by combining images of  
117 galaxies from the HRS. When we restrict the sample to the 45 galaxies with 
$R_{25} > $1.5\arcmin\ we can trace dust at 250\micron\ out to a radius of 1.75\,$R_{25}$ at $>$\,3$\sigma$.\\

\item We fit modified blackbody SEDs to the SPIRE and PACS (100--500\micron) radial profiles. 
We find the dust temperature declines from $\sim$25\,K in the centre of the galaxy to 
$\sim$15\,K at 1.0\,$R_{25}$ where it remains constant out to 1.75\,$R_{25}$.\\

\item We find the radial dust-mass distribution in a galaxy is consistent with 
an exponential profile out to a radius of 1.8\,$R_{25}$. The gradient of the surface-density profile is 
-1.7\,dex\,$R^{-1}_{25}$.\\

\item We created profiles of the stellar mass (traced by 3.4 or 3.6\micron) and SFR (UV \& Infrared) using images from
\Spitzer, GALEX and WISE. 
We found the stellar gradient is in good agreement with the dust gradient with
the dust gradient between 1\% and 10\% larger (in dex\,$R^{-1}_{25}$) depending on the dataset used. 
The SFR profile is significantly steeper than that of the dust with a gradient of $\sim$-2.6\,dex\,$R^{-1}_{25}$.\\

\item Our results show a much steeper gradient in the surface density of dust compared to metals (based on typical metallicity gradients 
for nearby galaxies), similar to the results for individual galaxies \citep[e.g.,][]{Mattsson2012b}. In closed-box chemical evolution models this is 
a signature that dust originates from grain growth in the ISM.\\

\item We use our results for the radial distribution of dust in galaxies to see if the large extended halos of dust around galaxies as described by \citet{Menard2010} could have a significant contamination caused by the clustering of galaxies and the dust emission within individual galaxies.
By using the number counts and angular correlation function from SDSS, we show that, assuming Milky Way type dust, the
dust within galaxy disks might explain the reddening results. Because of the large uncertainties in our analysis and the even larger
uncertainty of whether the dust in intergalactic space is like dust in galactic disks, we cannot rule out the 
the existence of intergalactic dust. We propose a simple method of 
testing whether there is actually intergalactic dust.

\end{enumerate}

Our results from a statistical stacking analysis with images from the \textit{Herschel Space Observatory} represent the most sensitive study of the extended 
dust emission around galaxies that will be possible until a new submillimetre space mission. It is likely to be impossible to improve on our results with
ground-based telescopes because of the fluctuating emission from the atmosphere, which makes it difficult to detect low surface-brightness emission from galaxies.

\section*{Acknowledgements}

We thank everyone involved with the {\it Herschel Space Observatory}.

SPIRE has been developed by a consortium of
institutes led by Cardiff University (UK) and including:
University of Lethbridge (Canada); NAOC (China); CEA, LAM (France); IFSI,
University of Padua (Italy); IAC (Spain); Stockholm Observatory (Sweden);
Imperial College London, RAL, UCL-MSSL, UKATC, University of Sussex (UK); and
Caltech, JPL, NHSC, University of Colorado (USA). This development has been
supported by national funding agencies: CSA (Canada); NAOC (China);
CEA, CNES, CNRS (France); ASI (Italy); MCINN (Spain); 
SNSB (Sweden); STFC, UKSA (UK); and NASA (USA). 

HIPE is a joint
development by the \Hersc\ Science Ground Segment Consortium,
consisting of ESA, the NASA \Hersc\ Science Center and the HIFI, PACS,
and SPIRE consortia. 

This research made use of Astropy, a community-developed core Python package for Astronomy (Astropy Collaboration, 2013).

HLG acknowledges support from the European Research Council (ERC) in the form of
Consolidator Grant {\sc CosmicDust} (ERC-2014-CoG-647939).




\bibliographystyle{mnras}
\bibliography{profile} 





%
%


\bsp	
\label{lastpage}
\end{document}